\documentclass[pss,fleqn]{w-art}
\usepackage{times}
\usepackage{w-thm}
\usepackage[]{epsfig,graphicx}
\begin{document}
\DOIsuffix{theDOIsuffix}
\Volume{XX}
\Issue{1}
\Month{01}
\Year{2004}
\pagespan{1}{}
\Receiveddate{9 September 2006}

\title[Thermoelectric effects in STM]{Thermoelectric effects in STM tunneling
       through a monoatomic chain}

\author[M. Krawiec]{M. Krawiec\footnote{Corresponding
     author: e-mail: {\sf krawiec@kft.umcs.lublin.pl}, Phone:
+48\,81\,5376146,
     Fax: +48\,81\,5376190}} 
\author[M. Jalochowski]{M. Ja\l ochowski} 
\address[]{Institute of Physics and Nanotechnology Center,
M. Curie-Sk\l{}odowska University,
 pl. M. Curie-Sk\l{}odowskiej 1, 20-031 Lublin, Poland}

\begin{abstract}
We study thermoelectric properties of the system composed of a monoatomic chain 
on a surface and additional electrode coupled to the chain, which can be an STM
tip. In particular, we are interested in thermopower, electric and thermal
conductance, Wiedemann-Franz relation and thermoelectric figure of merit, which
is a direct measure of the usefulness of the system for applications. We 
discuss the modifications of the STM wire topography due to temperature 
gradient between the electrodes. Finally, we also make connection to STM 
experiment, in which the thermopower has been directly measured, showing 
different structure, not visible in topography spectra.
\end{abstract}

\maketitle


\section{Introduction \label{intro}}

Since the invention of scanning tunneling microscopy (STM) \cite{Binnig} it
became possible to tailor and analyze small nanostructures on various 
conducting surfaces \cite{Briggs,Hofer}. As the STM is a real space technique
and is very sensitive to local atomic and electronic structures, it has been 
widely applied to study various surface reconstructions \cite{Hofer} and low
dimensional structures, like single atoms \cite{Briggs,Hofer}, islands
\cite{Jalochowski_1,MK_1} or one-dimensional monoatomic chains
\cite{Himpsel}-\cite{MK_2}. While the STM is used to characterize various
structures measuring topography or current-voltage characteristics, it is also
suitable to get information on thermal properties of the system. For example, 
local thermopower S({\bf r}) of various surfaces has been measured with STM
\cite{Weaver,Williams}. Thus the topography of the surface can be supplemented 
by its 'thermal' image. It turns out that the thermal images show features that 
are not accessible in the conventional STM topographic images. In particular, 
the maxima of the thermopower in the thermal image of the surface do not 
overlap with the tunneling current maxima \cite{Weaver,Williams}. The local 
thermopower of the surfaces observed in STM experiments \cite{Weaver,Williams} 
has been also studied theoretically \cite{Stovneng} within the Tersoff-Hamann 
approach \cite{Tersoff}. It was found that the magnitude of the thermopower 
depends on the logarithmic derivative of the local surface density of states at 
the Fermi level and does not show exponential dependence on tip-surface 
distance, in contrast to the tunneling current.

The above examples show that thermoelectric properties are the source of 
information complementary to that obtained from other transport 
characteristics. In bulk systems, transport in the presence of electrical and
thermal gradients is a well studied phenomenon. Recently, it became possible to
study those effects, both experimentally and theoretically, in atomic-scale 
systems, like quantum point contacts \cite{vanHouten,Molenkamp}, quantum dots 
\cite{Beenakker}-\cite{MK_3}, quantum dot superlattices \cite{Heremans}, 
quantum wires with end atoms coupled to external leads \cite{Fazio}, carbon 
nanotubes \cite{Lin,Pop} or correlated multilayered nanostructures 
\cite{Freericks}.

It is the purpose of the present work to study thermoelectric properties of 
quantum wire in STM geometry. We shall concentrate on the electric and thermal 
conductance, thermopower and related quantities, like thermoelectric figure of 
merit which is a direct measure of the usefulness of the system for 
applications and Wiedemann-Franz ratio which signals breakdown of the Fermi 
liquid state. We also discuss the modifications of the STM wire topography due 
to temperature gradient between the electrodes. To explore electric and thermal 
properties of the system we propose a model of tunneling between STM tip and 
surface and use equations of motion technique for Green functions. Rest of the 
paper is organized as follows: in Sec. \ref{model} we introduce our model based 
on tight binding model and discuss some aspects of our procedure, and the 
results of our calculations are presented in Sec. \ref{results}. We end up with 
summary and conclusions.


\section{The model \label{model}}

Our model system is described by the Hamiltonian
\begin{eqnarray}
H = \sum_{\lambda \in \{\mathrm{t,s}\}{\bf k}\sigma} \epsilon_{\lambda {\bf k}} 
c^+_{\lambda {\bf k} \sigma} c_{\lambda {\bf k} \sigma} +
\sum_{\sigma} \mathrm{\varepsilon}_0 c^+_{0\sigma} c_{0\sigma} +
\sum_{i\sigma} \mathrm{\varepsilon}_{\mathrm{w}} c^+_{i\sigma} c_{i\sigma} + 
\sum_{ij\sigma} \left( \mathrm{t}_{\mathrm{w}} c^+_{i\sigma} c_{j\sigma} + 
H.c. \right) 
\nonumber \\
+ \sum_{{\bf k}\sigma} \left( \mathrm{V}_{\mathrm{t}} 
c^+_{{\mathrm{t}} {\bf k} \sigma} c_{0\sigma} + H. c. 
\right) +
\sum_{i {\bf k}\sigma} \left( \mathrm{V}_{i\mathrm{s}} e^{\imath {\bf k R}_i} 
c^+_{{\mathrm{s}} {\bf k} \sigma} c_{i\sigma} + H. c. \right) +
\sum_{i\sigma} \left( \mathrm{t}_{i0} c^+_{0\sigma} c_{i\sigma} + H. c. 
\right),
\label{hamilt}
\end{eqnarray}
and consists of N-atom metallic wire with atomic energies 
$\mathrm{\varepsilon}_{\mathrm{w}}$ and hopping parameter t$_{\mathrm{w}}$ 
between neighboring wire atoms. The wire is connected to the surface via 
parameters V$_{i\mathrm{s}}$, which we treat as a reservoir for electrons with 
energies $\mathrm{\epsilon}_{\mathrm{s} {\bf k}}$. STM tip is modeled by single 
atom with energy $\mathrm{\varepsilon}_0$ attached to another reservoir (with 
electron energies $\mathrm{\epsilon}_{\mathrm{t} {\bf k}}$) via coupling 
V$_\mathrm{t}$. Tunneling between STM tip and the wire atoms is described by 
tunneling matrix element t$_{i0}$. As usually, $c^+_{\lambda}$ ($c_{\lambda}$) 
stands for creation (annihilation) electron operator in STM lead 
($\lambda = \mathrm{t}$), tip atom ($\lambda = 0$), $i$th wire atom 
($\lambda = i$) and surface ($\lambda = \mathrm{s}$). Schematic view of our 
model system is shown in Fig. 1.
\begin{figure}[h]
\begin{center}
\includegraphics[width=0.21\textwidth]{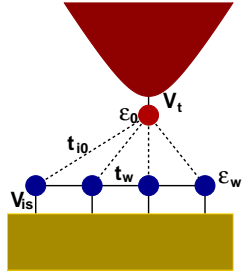}
\caption{Schematic view of our model system composed of N-atom wire on surface 
         and STM tip.}
\end{center}
\label{Fig1}
\end{figure}

In order to calculate tunneling current $J_e$ and thermal flux $J_Q$ flowing
from the STM electrode to the rest of the system we follow standard derivation 
\cite{MK_3} and get
\begin{eqnarray}
J_e = \frac{e}{h} \int d\omega \mathcal{T}(\omega)
\left[f(\omega - \mu_{\mathrm{t}}) - f(\omega - \mu_{\mathrm{s}}) \right] ,
\label{e_current}
\end{eqnarray}
\begin{eqnarray}
J_Q = \frac{1}{h} \int d\omega \mathcal{T}(\omega)
(\omega - \mu_{\mathrm{t}}) 
\left[f(\omega - \mu_{\mathrm{t}}) - f(\omega - \mu_{\mathrm{s}}) \right] , 
\label{Q_current}
\end{eqnarray}
where 
$\mathcal{T}(\omega) = 
2 \Gamma^{\mathrm{t}}(\omega) \sum_{ij} \Gamma^{\mathrm{s}}_{ij}(\omega) 
G^a_{j0}(\omega) G^r_{0j}(\omega)$ is the transmittance of the system, 
$f(\omega)$ is the Fermi distribution function, and $\mu_{\mathrm{t}}$ 
($\mu_{\mathrm{s}}$) is the chemical potential of the tip (surface) electrode. 
$G^{r(a)}_{j0}(\omega)$ is the matrix element (connecting the tip atom with the 
$j$th wire atom) of the retarded (advanced) Green function $\hat G(\omega)$, 
the solution of the equation 
$(\omega \hat 1 - \hat H) \hat G(\omega) = \hat 1$. The parameter 
$\Gamma^{\mathrm{t}}(\omega) = \sum_{\bf k} |{\mathrm{V}}_{\mathrm{t}}|^2 
\delta(\omega - \epsilon_{{\mathrm{t}}{\bf k}})$ denotes strength of the 
coupling between the tip atom and STM electrode, while 
$\Gamma^{\mathrm{s}}_{ij}(\omega) = \sum_{\bf k} |{\mathrm{V}}_{\mathrm{s}}|^2 
e^{\imath {\bf k}({\bf R}_i - {\bf R}_j)} 
\delta(\omega - \epsilon_{{\mathrm{t}}{\bf k}}) 
= \Gamma^{\mathrm{s}}(\omega) \mathrm{sin}(k_F a (i-j))/(k_F a (i-j))$ is the 
coupling between wire atoms $i$ and $j$ via the surface. $k_F$ is the Fermi 
wave vector of the surface electrode, and $a$ is a distance between neighboring 
wire atoms. In typical metals $k_F a$ is of order of $4 - 5$.

For small bias voltages 
$eV = \mu_{\mathrm{t}} - \mu_{\mathrm{s}} \rightarrow 0$ and small 
temperature gradients 
$\delta {\mathrm{T}} = {\mathrm{T}}_{\mathrm{t}} - {\mathrm{T}}_{\mathrm{s}} 
\rightarrow 0$ one defines conductance 
${\mathrm{G}} = -(e^2/{\mathrm{T}}) L_{11}$, thermopower is given in the form 
${\mathrm{S}} = -(1/e{\mathrm{T}}) (L_{12}/L_{11})$, and thermal conductance 
$\kappa = (1/{\mathrm{T}}^2) (L_{22} - L^2_{12}/L_{11})$. The linear response 
coefficients read
\begin{eqnarray}
L_{11} = \frac{\mathrm{T}}{h} \int d\omega \mathcal{T}(\omega) 
\left(\frac{\partial f(\omega)}{\partial \mu} \right)_{\mathrm{T}} , 
\label{L11}
\end{eqnarray}
\begin{eqnarray}
L_{12} = \frac{{\mathrm{T}}^2}{h} \int d\omega \mathcal{T}(\omega) 
\left(\frac{\partial f(\omega)}{\partial {\mathrm{T}}} \right)_{\mu} , 
\label{L12}
\end{eqnarray}
\begin{eqnarray}
L_{22} = \frac{{\mathrm{T}}^2}{h} \int d\omega \mathcal{T}(\omega) 
(\omega - \mu_{\mathrm{t}}) 
\left(\frac{\partial f(\omega)}{\partial {\mathrm{T}}} \right)_{\mu} .
\label{L22}
\end{eqnarray}
%


\section{Results and discussion \label{results}}

Before the presentation of numerical results, we would like to comment on 
choice of the model parameters used in the present work. In numerical 
calculations we have assumed equal and energy independent coupling parameters 
($\Gamma^{\mathrm{t(s)}}(\omega) = \Gamma^{\mathrm{t(s)}}$) and chosen 
$\Gamma^{\mathrm{s}}= \Gamma^{\mathrm{t}} = \Gamma$ as an energy unit. The 
other parameters have been chosen in order to satisfy realistic situation in 
experiments. The hopping integral along the wire is 
${\mathrm{t}}_{\mathrm{w}} = 2$, the parameter connecting tip with underneath 
wire atom ${\mathrm{t}}_{i0} = 0.1$ and STM tip and wire atomic energies 
$\varepsilon_0 = \varepsilon_{\mathrm{w}} = 0$. For example, taking 
$\Gamma = 0.05$ eV, we get ${\mathrm{t}}_{\mathrm{w}} = 0.1$ eV and 
${\mathrm{t}}_{i0} = 0.005$ eV. Such a value of ${\mathrm{t}}_{i0}$ gives 
tip-surface distance ${\mathrm{z}}_{i0} = 6 \; \mathrm{\AA}$. Also temperature 
is measured in units of $\Gamma$ ($k_B = 1$) and for example, T = 1 corresponds 
to $\approx 500$ K. 

Figure 2 shows conductance G (top panels), thermal conductance $\kappa$ (middle 
panels) and thermopower S (bottom panels) of quantum wire.
\begin{figure}[h]
\begin{center}
\includegraphics[width=0.51\textwidth]{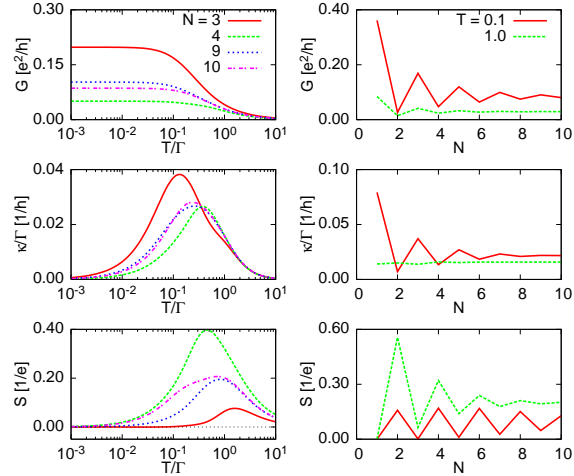}
\caption{Conductance (top panels), thermal conductance (middle panels) and
thermopower (bottom panels) as a function of temperature (left panels) and as a
function of number of atoms in the wire (right panels). All the curves
correspond to situation when STM tip is placed above first wire atom, and the
tip surface distance is equal to $6 \; \mathrm{\AA}$.}
\end{center}
\label{Fig2}
\end{figure}
The temperature dependencies of those quantities for wires consisted of 
different number of atoms (indicated in the figure) are shown in left panels, 
while in right panels the same quantities are plotted as a function of number 
of atoms in the wire at different temperatures. In all the cases the STM tip is
above first wire atom, and the tip-surface distance is equal to 
$6 \; \mathrm{\AA}$. 

The conductance G shows rather an expected behavior, i.e. at high T it goes 
like $\mathrm{T}^{-1}$, while in the low temperature regime remains constant. 
Note, that it never reaches unitary limit ($2e^2/h$) due to large tip surface 
distance. However there are differences in maximal values of G for different
number of atoms N in the wire. To be more precise, the conductance shows 
even-odd oscillations with increasing N (see right top panel). Such a 
behavior is similar to that observed in the transport along wires 
\cite{Agrait}. Similar oscillations shows thermal conductance $\kappa$, 
although they are less pronounced at higher temperatures (see right middle 
panel). However, it has different temperature dependence, it goes linearly with 
T at low temperatures, while at high T shows $\mathrm{T}^{-2}$ behavior. The 
thermopower S goes linearly to zero with T and behaves like $\mathrm{T}^{-1}$ 
in high temperature regime. It is almost always positive in the whole range 
temperatures, suggesting an electron nature of transport. Interestingly, it 
shows also even-odd oscillations with number of atoms in a wire but with 
opposite phase to G and $\kappa$, i.e. while G and $\kappa$ show maxima for the
wires composed of odd number of atoms, S has maximal values for even N. 
This can be easily understood, as the conductance G is sensitive to density 
of states (DOS) at the Fermi level $E_F$, while S reflects the curvature of 
DOS around it. In odd atom wire the DOS is large at $E_F$ and has small 
curvature, while in even atom wire the situation is opposite. Note that the 
presence of the surface introduces asymmetry to the DOS around the Fermi 
energy, even if we assume wire energies to coincide with $E_F$ 
\cite{Newns,MK_2}.

Topography of the wire consisted of N = 19 atoms at $eV = 0.1$ and 
$J_e = 10^{-5}$ (V = 0.5 mV and $J_e = 0.2$ nA) is shown in Fig. 3. 
\begin{figure}[h]
\begin{center}
\includegraphics[width=0.51\textwidth]{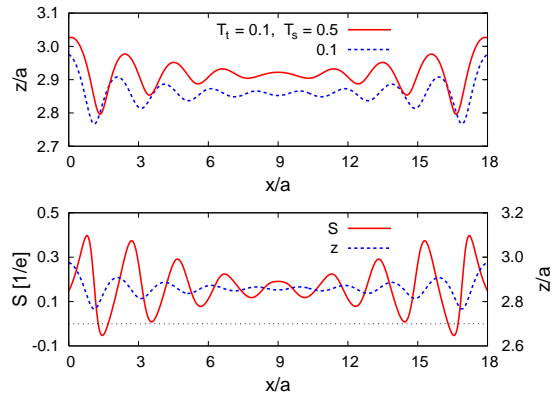}
\caption{Topography of the wire $\mathrm{z}(\mathrm{x})$ at $eV = 0.1$ and 
         $J_e = 10^{-5}$ (V = 0.5 mV and $J_e = 0.2$ nA) for the STM tip 
	 temperature $\mathrm{T}_{\mathrm{t}} = 0.1$ and different temperatures 
	 of the surface electrode $\mathrm{T}_{\mathrm{s}}$ (top panel). 
	 Corresponding thermopower S(x) is shown in the bottom panel. For 
	 comparison z(x) is also displayed. x and z are in units of the 
	 distance between neighboring wire atoms $a$.}
\end{center}
\label{Fig3}
\end{figure}
Dashed line corresponds to the situation when STM tip and the surface
temperatures are equal 
($\mathrm{T}_{\mathrm{t}} = \mathrm{T}_{\mathrm{s}} = 0.1$ (50 K)), while the 
solid one to the situation when surface is at elevated temperature 
$\mathrm{T}_{\mathrm{s}} = 0.5$ (250 K). The reverse of the topography is 
clearly seen due to the temperature difference between STM tip and the surface. 
To be convinced that this is really the case, we have also shown thermopower 
S(x) (solid line in the bottom panel). Note that the maxima of S(x) do not 
coincide with the topography maxima calculated for 
$\mathrm{T}_{\mathrm{t}} = \mathrm{T}_{\mathrm{s}}$ (dashed line in the bottom 
panel) but they do with those obtained at different STM tip and surface 
temperatures (dashed line in the top panel). Thus indeed the reverse of the 
topography is caused by the temperature gradient between STM tip and surface. 
Similar differences between the topography and thermal images of the surface 
have been observed experimentally with STM \cite{Williams}.

The Wiedemann-Franz (WF) relation $\kappa/{\mathrm{TG}} = \pi^2/3 e^2$ 
describes transport in Fermi liquid bulk metals and in general is not obeyed in 
mesoscopic systems. In our system there are no Coulomb interactions and this 
relation is fulfilled at low temperatures, thus indicating Fermi liquid ground 
state. This can be also understood from temperature behavior of electric and 
thermal conductances, as $\kappa$ decreases linearly with T, while G remains 
constant. At high temperatures the WF relation is violated due to large 
fluctuations of G, which are of order of $2 e^2/h$ \cite{Vavilov}. 

Finally, we would like to comment on usefulness of such systems for potential
applications, like thermoelectric power generators or cooling systems
\cite{Heremans}. 
A direct measure of it is thermoelectric figure of merit 
$\mathrm{Z} = \mathrm{S}^2 \mathrm{G}/\kappa$. For simple systems it is 
inversely proportional to operation temperature, thus conveniently consider ZT, 
which numerical value is a measure of the system performance. In our system the 
value of ZT is always smaller than 1, indicating limited practical 
applicability. Also the thermopower, which is of order of $10^{-2}$ mV/K, leads 
to similar conclusions.


\section{Conclusions \label{concl}}

In conclusion we have studied thermal properties of monoatomic quantum wire in
STM geometry, and show differences between topography and thermal images of the
wire. In particular, the maxima of calculated thermopower along the wire do not
coincide with those of topography spectra due to the fact that tunneling current
is sensitive to the density of states, while thermopower is sensitive to the
slope of the density of states at the Fermi energy. We discussed also 
Wiedemann-Franz relation and found that is fulfilled at low temperatures 
indicating Fermi liquid ground state.


\begin{acknowledgement}
This work has been supported by the Grant No 1 P03B 004 28 of the Polish
Committee of Scientific Research.
\end{acknowledgement}


\end{document}